\documentstyle[12pt]{article}

\def\mypagenumber{1}
\def\mydate{August 22, 1996}
\def\myend{\end{document}}

\def\Journal#1#2#3#4{{#1}{\bf #2} (#4) #3}

\def\NPB{{\em Nucl.\ Phys.} B}
\def\PLB{{\em Phys.\ Lett.} B}
\def\PRL{\em Phys.\ Rev.\ Lett. }
\def\PRB{{\em Phys.\ Rev.} B}
\def\PRD{{\em Phys.\ Rev.} D}
\def\AP{{\em Ann.\ Phys.\ (N.Y.)} }
\def\RMP{{\em Rev.\ Mod.\ Phys.} }

\def\CMP{\em Comm.\ Math.\ Phys. }
\def\MPLA{{\em Mod.\ Phys.\ Lett.} A}
\def\IJMPB{{\em Int.\ J.\ Mod.\ Phys.} B}

\normalsize

\newcounter{sxn}

\newcounter{axn}

\date{}

\newdimen\mybaselineskip
\newcommand{\beeq}{\begin{equation}}
\newcommand{\eneq}{\end{equation}}
\newcommand{\beqn}{\begin{eqnarray}}
\newcommand{\eeqn}{\end{eqnarray}}
\newcommand{\bpic}{\begin{picture}}
\newcommand{\epic}{\end{picture}}

\def\dd{\partial}
\def\la{\raise.16ex\hbox{$\langle$} \, }
\def\ra{\, \raise.16ex\hbox{$\rangle$} }
\def\go{\rightarrow}

\def\next{{~~~,~~~}}

\def\psibar{ \psi \kern-.65em\raise.6em\hbox{$-$} }
\def\mbar{ m \kern-.78em\raise.4em\hbox{$-$}\lower.4em\hbox{} }

\def\Abar{{\overline A}}
\def\Bbar{{\overline B}}
\def\Cbar{{\overline C}}
\def\Dbar{{\overline D}}

\def\tkappa{{\tilde\kappa}}
\def\tM{{\tilde M}}

\def\L{ {\cal L} }
\def\ep{\epsilon}
\def\eps{\varepsilon^{\mu\nu\rho}}

\def\hg{{\hat g}}
\def\hp{{\hat p}}

\def\eff{{\rm eff}}

\def\n@space{\nulldelimiterspace=0pt \mathsurround=0pt }
\def\huge#1{{\hbox{$\left#1\vbox to 20.5pt{}\right.\n@space$}}}

\def\myskip{\noalign{\kern 10pt}}
\def\myeqspace{\noalign{\kern 10pt}}
\def\crn{\cr\myeqspace}

\def\boxit#1{$\vcenter{\hrule\hbox{\vrule\kern3pt
    \vbox{\kern3pt\hbox{#1}\kern3pt}\kern3pt\vrule}\hrule}$}
\def\bigbox#1{$\vcenter{\hrule\hbox{\vrule\kern5pt
     \vbox{\kern5pt\hbox{#1}\kern5pt}\kern5pt\vrule}\hrule}$}

\def\ignore#1{{}}

\begin{document}

\bibliographystyle{unsrt}
\footskip 1.0cm

\thispagestyle{empty}
\setcounter{page}{\mypagenumber}

{\baselineskip=10pt \parindent=0pt \small
Revised.  \mydate 
\hfill \hbox{\vtop{\hsize=3.2cm   UMN-TH-1505/96\\  hep-th/9607233\\}}
%\vspace{6mm}
%\rightline{\mydate}
}

\vspace*{25mm}

\centerline {\Large\bf  Spontaneous Symmetry Breaking at Two Loop}
\vspace*{8mm}
\centerline{\Large\bf  in 3-d Massless Scalar Electrodynamics}
\vspace*{30mm}
\centerline{\large Pang-Ning Tan\footnote{e-mail:~ tan@mnhepo.hep.umn.edu},
 Bayram Tekin\footnote{e-mail:~ tekin@mnhepo.hep.umn.edu},
and Yutaka Hosotani\footnote{e-mail:~ yutaka@mnhepw.hep.umn.edu}}

\vspace*{10mm}

\baselineskip=15pt
\centerline {\it School of Physics and Astronomy, University
       of  Minnesota}
\centerline {\it Minneapolis, Minnesota 55455, U.S.A.}

\vspace*{25mm}
%\baselinestretch{2.0}

%\normalsize

\begin{abstract}
\baselineskip=18pt
In three dimensional Maxwell-Chern-Simons massless scalar electrodynamics
with  $ \phi^6$ coupling,  the $U(1)$ symmetry is  spontaneously
broken at two loop order
regardless of the presence or absence of the Maxwell term. 
Dimensional transmutation takes place in pure Chern-Simons scalar
electrodynamics. The beta function for the $\phi^6$ coupling is independent of
gauge  couplings.
\end{abstract}

\vspace*{5mm}

%\end{titlepage}
 
\newpage

%\setcounter{page}{1}

%\textheight=20cm
%\headsep=0.75cm
%\vsize=20cm

%%%%%%%%%%%%%%%%%%%%%%%%%%%%%%%%%%%%%%%%%%%%%%%%%%%%%%%%%%%%%%%%%
\normalsize
\baselineskip=22pt plus 1pt minus 1pt
\parindent=25pt
%\vspace*{5mm}

%\sxn{Introduction}

We evaluate the effective potential of massless scalar fields in
three-dimensional $U(1)$ gauge theory to the two loop order to 
examine whether or not the $U(1)$ symmetry is spontaneously broken
by radiative corrections.  We shall obtain  the effective potential 
in a closed form and demonstrate that the $U(1)$ symmetry
is spontaneously broken when the Chern-Simons term is present for  
gauge fields.

Three-dimensional gauge field theories are relevant to physics in many
respects.  Nonrelativistic Maxwell-Chern-Simons theory is a useful 
tool to describe condensed matter systems such as the quantum Hall 
effect \cite{QHE} and 
superconductivity.\cite{super}
It provides  field theory description of fractional
statistics or anyons.\cite{anyon}

Three-dimensional gauge theories also serve as effective theories of
 four dimensional  theories at high temperatures.\cite{Linde,Kajantie}
Although  power-like infrared divergences could make 
perturbation theory at high temperature in four dimensions unreliable,
perturbation theory becomes safe at small momenta or fields
once a theory is dimensionally reduced to 2+1 dimensions. 
 Maxwell-Chern-Simons
theory appears as an effective theory of QCD and the standard model of
electroweak  interactions.

Further Maxwell-Chern-Simons theory provides a unique testing ground for new 
theoretical ideas.  A photon acquires a topological mass.\cite{Deser,Schon} 
There appear vortex soliton solutions with both magnetic flux and charge.
 Multi-soliton solutions are also available.\cite{vortex} Supersymmetric
generalization exists.  In a theory with Dirac fermions a  magnetic field can be
dynamically generated so that the Lorentz invariance is spontaneously
broken.\cite{Hosotani}  Pure non-Abelian Chern-Simons theory defines a
topological field theory.  It plays an important role in the knot
theory.\cite{Witten}

In this paper we concentrate on another aspect of Maxwell-Chern-Simons theory,
namely the spontaneous breaking of the $U(1)$ gauge symmetry by radiative
corrections.\cite{CW,Bardeen,MCS}  In a theory with  massless  charged scalar
fields,  radiative  corrections control whether  or not the scalar fields
develop a non-vanishing expectation  value.  The limiting case is of special
interest, in which the Maxwell term is absent for gauge fields and the scalar
fields self-interact only through  $\phi^6$ coupling.   As the theory 
contains no dimensional parameter,  the question is now phrased as whether or
not  dimensional transmutation takes  place in three dimensional gauge theory.

In the literature the effective potential has been evaluated at one loop,
with sometimes conflicting conclusions.
We shall see below that one has to go beyond one loop, i.e. that two loop
corrections are decisive to determine the symmetry of the vacuum.  We evaluate
the effective  potential when gauge fields have both Maxwell and Chern-Simons
terms. We show that the limit of the vanishing Maxwell term is well defined after
renormalization.

A few comments are in order.   A Chern-Simons term is induced by 
Dirac fermions at one loop.\cite{DiracCS}    In non-Abelian theory the
Chern-Simons coefficient is quantized.\cite{Deser}  Non-Abelian gauge fields
themselves induce a  Chern-Simons term at one loop.\cite{Pisarski}
Pure non-Abelian Chern-Simons theory is ultraviolet finite.\cite{Martin}
It has also been argued that a Chern-Simons term is induced 
by scalar fields with spontaneous symmetry breaking.\cite{scalarCS}

Secondly it is safer to start with both Maxwell and Chern-Simons 
terms in order to perform computations beyond one loop.  The presence 
or absence of the Maxwell term drastically changes the ultra-violet behavior 
of the gauge field propagator.  As we shall see below, the photon 
propagator in the dimensional regularization scheme becomes problematic
if the Maxwell term is absent.  The limit of the vanishing renormalized 
coefficient of the Maxwell term should be taken at the end of computations.

Long time ago Coleman and Weinberg showed that in four-dimensional massless
scalar electrodynamics, spontaneous symmetry breaking  is induced by radiative
corrections.\cite{CW}  They calculated one loop corrections to the effective
potential and showed that the scalar field develops a nonvanishing vacuum
expectation value.  In our case two loop computation is necessary to see it.

The most general, renormalizable Lagrangian density containing a
complex scalar field ($=\phi_1+i\phi_2$) and an Abelian  gauge field is 
\beqn
\L &=& - {a\over 4} F_{\mu\nu} F^{\mu\nu}
- {\kappa\over 2} \ep^{\mu\nu\rho} A_\mu \dd_\nu A_\rho
+ \L_{\rm g.f.} + \L_{\rm F.P.} \cr
\noalign{\kern 5pt}
&& +{1\over 2} (\dd_\mu\phi_1 - eA_\mu \phi_2)^2
 +{1\over 2} (\dd_\mu\phi_2 + eA_\mu \phi_1)^2 \cr
\noalign{\kern 5pt}
&&-{m^2\over 2} (\phi_1^2 +\phi_2^2) 
- {\lambda\over 4!}  (\phi_1^2 +\phi_2^2)^2 
-{\nu\over 6!}  (\phi_1^2 +\phi_2^2)^3 ~~. 
  \label{Lagrangian1}
\eeqn
where the gauge fixing term (in $R_\xi $ gauge) along with the 
Faddeev-Popov ghost term are 
\beqn
\L_{\rm g.f.}~ &=& - {1\over 2\xi} (\dd_\mu A^\mu - \xi ev \phi_2)^2 \cr
\noalign{\kern 5pt}
\L_{\rm F.P.} &=& - c^\dagger \, ( \dd^2 + \xi e^2 v \phi_1) \, c ~~.
\label{R-gauge}
\eeqn
One of the coupling constants $a, \kappa, e,$ is redundant, as it can
be absorbed in the redefinition of $A_\mu$.  For instance, if one expresses
$\L$ in terms of  $A'_\mu=eA_\mu$, then we have $a'=a/e^2$, $\kappa'=\kappa/e^2$
and $e'=1$.  The Coleman-Weinberg limit is defined by the vanishing of
renormalized $a$, $m^2$, and $\lambda$.

We choose the 
vacuum expectation values to be $\la \phi_1 \ra = v$ and $\la \phi_2
\ra =0$. 
After shifting the field $\phi_1 \go v + \phi_1$,  the quadratic part is
\beqn
\L^{(0)} &=& 
{1\over 2} A_\mu ~ K^{\mu\nu} ~  A_\nu 
- c^\dagger \, ( \dd^2 + \xi e^2 v^2) \, c
 - {1\over 2} \phi_1 (\dd^2 + m_1^2 ) \phi_1 
 - {1\over 2} \phi_2 (\dd^2 + m_2^2 ) \phi_2  \cr
\noalign{\kern 8pt}
K^{\mu\nu}  &=& 
\bigg\{  a  \dd^2  + (e v)^2 \bigg\}g^{\mu\nu} 
  - \bigg( a - {1\over \xi} \bigg) \dd^\mu\dd^\nu  
+ \kappa \, \ep^{\mu\nu\rho} \dd_\rho 
\label{Lagrangian2}
\eeqn
where
\beqn
m_1^2[v] &=& m^2 + {\lambda\over 2} v^2   + {\nu\over 24} v^4\cr
\noalign{\kern 8pt}
m_2^2[v] &=&  m^2 + {\lambda\over 6} v^2 +  {\nu\over 120} v^4
   + \xi (ev)^2 ~~.
\label{mass1}
\eeqn

Throughout the paper we employ the dimensional regularization method.
Evaluation of radiative corrections beyond one loop is simplified in the
dimensional regularization.  However, in the theory at hand we need to define
the  antisymmetric tensor $\eps$ in $n$ dimensions.  There is no natural
definition which preserves the Lorentz invariance.
We adopt the definition given by\cite{tHooft}
\beeq
\eps = \cases{
\pm 1 &if $(\mu,\nu,\rho)$= permutation of (0,1,2)\cr
0&otherwise.\cr}
\label{epsilon-tensor}
\eneq
Similarly we  define three-dimensional metric $\hg^{\mu\nu}$:
\beeq
\hg^{\mu\nu} = \cases{ +1&for $\mu=\nu=0$\cr
                       -1&for $\mu=\nu=1,2$\cr
                        0&otherwise.\cr}
\label{3dmetric}
\eneq
This definition of $\eps$ has been commonly employed  in the investigation of 
higher order corrections in pure non-Abelian Chern-Simons 
theory.\cite{Martin}
Of course one needs to ensure that this definition would not spoil 
the gauge invariance.  Although full investigation is reserved in future
publication, we have confirmed that the dimensional regularization,
cutoff method, zeta function regularization, all give the the same result at one
loop.  It also has been shown in \cite{Chen} that in pure non-Abelian
Chern-Simons theory Slavnov-Taylor identities are satisfied up to
three loop order with this definition of $\eps$ in the dimensional
regularization scheme.

The gauge propagator can be found easily. In the Landau gauge ($\xi=0$)
\beeq
K^{-1}_{\nu\lambda}
= -{1\over d(p^2)} \bigg( g_{\nu\lambda} 
 -  {p_\nu p_\lambda \over p^2 }  \bigg)  - {\kappa^2 \hp^2\over d(p^2) [d(p^2)^2  - \kappa^2 \hp^2]} \,
  \bigg(\hg_{\nu\lambda} - {\hp_\nu \hp_\lambda\over \hp^2} \bigg) 
    + {i\kappa \ep_{\nu\lambda\rho} p^\rho
   \over d(p^2)^2 - \kappa^2 \hp^2} 
\label{Kinverse1}
\eneq
where $d(p^2) = ap^2 - (ev)^2$.  Here 
$\hat p^\mu = \hat g^{\mu\nu} p_\nu = (p^0, p^1, p^2, 0 , \cdots, 0)$.
The propagator is rearranged in the form
\beqn
K^{-1}_{\mu\nu}
&=& - {1\over a} \bigg\{ {1\over {m_+ + m_-}} \bigg( 
      {1\over m_+}{1\over{p^2 - m_+^2}} + {1\over m_-}{1\over{p^2 - m_-^2}}
      \bigg) - {1\over m_3^2}{1\over{p^2}} \bigg\}
      (\hg_{\mu\nu}\hp^2 - \hp_\mu \hp_\nu) \cr
\myskip
&&  + {1\over a^2}{1\over m_+^2 - m_-^2} \bigg( {1\over p^2 - m_+^2}
    - {1\over p^2 - m_-^2} \bigg) i\kappa \ep_{\mu\nu\rho} p^\rho \cr
\myskip
&&  - {1\over a}{1\over m_3^2}  \bigg( {1\over p^2 - m_3^2}
    - {1\over p^2} \bigg) \bigg( (g_{\mu\nu}p^2 - p_\mu p_\nu) -
    (\hg_{\mu\nu}\hp^2 - \hp_\mu \hp_\nu) \bigg) \cr
\myskip
&&  + {\kappa^2 (p^2 - \hp^2)\over (d^2-\kappa^2p^2)(d^2-\kappa^2 \hp^2)} 
    \bigg\{  {\kappa^2\over d} (\hg_{\mu\nu}\hp^2 - \hp_\mu \hp_\nu)
    - i\kappa \ep_{\mu\nu\rho} p^\rho \bigg\}~~~.
\label{Kinverse2}
\eeqn
Here 
\beqn
m_{\pm}(v) &=& {1\over 2} \, \left\{ \sqrt{ {\kappa^2\over a^2} 
        + {4(ev)^2\over a}} \pm {|\kappa|\over a} \right\}  \cr
m_3^2(v) &=& m_+ m_- = {(ev)^2\over a}
\label{mass2}
\eeqn
There are two propagating modes with masses $m_+$ and $m_-$ which are
combinations of Higgs and Chern-Simons masses.  The massless pole represents
a gauge degree of freedom, whereas $m_3$ is the mass of photon in 
extra-dimensional space.  

Notice that the propagator (\ref{Kinverse1}) behaves as $1/p^2$ for large
$p^2$, provided $a \not= 0$.  If $a=0$ and $v\not= 0$, the extra-dimensional
components of the propagator  behave badly ($\sim$O(1)) for large $p^2$.
In intermediate stage of calculations we need to keep $a$ nonvanishing.
The $a \go 0$ limit must be taken at the end after renormalization.

Before  proceeding to two loop calculations of the effective potential for the 
Maxwell-Chern-Simons theory, it is instructive to look at the pure scalar 
part of the theory first. One loop calculations yield the following result;
\beeq
V_\eff^{(1)} =
-{\hbar\over 12\pi} \, \mu^{n-3} \Big\{ m_1(v)^3 + m_2(v)^3  \Big\}
\eneq
where $m_j(v)$'s are given in (\ref{mass1}) with $\xi=0$.

A typical two-loop integral in the Euclidean space is \cite{Kajantie}
\beqn
I( m_1,m_2,m_3;n) &\equiv& \int \frac{d^nqd^nk}{(2\pi)^{2n}}
   \frac{1}{\left[ (q+k)^2+m_1^2\right] (q^2+m_2^2)(k^2+m_3^2)}\crn
&=& I(m_2, m_1,m_3;n) \quad {\rm etc.} \crn
&=&
{\mu^{2(n-3)}\over 32 \pi^2}  \bigg\{ - {1\over n-3} - \gamma_E + 1 
- \ln {(m_1+m_2+m_3)^2\over 4\pi \mu^2} \bigg\}
\label{Ifunction}
\eeqn
In the pure scalar theory
\beqn
&&\L_{(\rm cubic)} = - \beta_1 \phi_1^3 - \beta_2 \phi_1 \phi_2^2 \cr 
\myskip
&&\L_{(\rm quartic)} = - \alpha_1 \phi_1^4  - \alpha_2 \phi_2^4  
             - \alpha_3 \phi_1^2\phi_2^2  \cr 
%\myskip
%&& {\rm where} \cr
\myskip
&&\beta_1 = {\lambda\over 3!} v + {\nu\over 36} v^3 \next
  \beta_2 = {\lambda\over 3!} v + {\nu\over 60} v^3 \cr 
\myskip
&&\alpha_1 = {\lambda\over 4!} + {\nu\over 2\cdot 4!} v^2 \next
  \alpha_2 = {\lambda\over 4!} + {\nu\over 2\cdot 5!} v^2 \next
  \alpha_3 = {2\lambda\over 4!} + {3\nu\over 5!} v^2 ~~.
\eeqn
Two loop contributions are given by
\beqn
V_\eff^{(2)} &=&
-\hbar^2 \bigg\{ 3 \beta_1^2 \, I(m_1,m_1,m_1)
   + \beta_2^2 \, I(m_1,m_2,m_2)  \bigg\} \cr \myskip
&& + {\hbar^2\over 16\pi^2} \, \mu^{2(n-3)}   \bigg\{
3 \alpha_1 m_1^2 + 3 \alpha_2 m_2^2 + \alpha_3 m_1 m_2 \bigg\}
\eeqn
The total effective potential is 
$V_\eff = V^{(0)} + V^{(1)} + V^{(2)} + V^{\rm counter ~terms}$.

The existence of symmetry breaking depends on renormalized $m^2$, $\lambda$,
and $\nu$.  Several authors have considered cases where
the renormalized $m^2, \lambda > 0$  but $\nu=0$,  and found that
there is symmetry breaking at one loop \cite{MCS}  
in contrast to 
 the (3+1) dimensional pure scalar theory where there is no
symmetry breaking \cite{CW}.

Let us focus on a special case $m^2=\lambda=0$.  
We impose the following renormalization conditions at $n=3$:
\beqn
&&{\dd^2 V_\eff\over \dd v^2} \Bigg|_{v=0} = m^2 = 0 \cr \myskip
&&{\dd^4 V_\eff\over \dd v^4} \Bigg|_{v=0} = \lambda = 0 \cr \myskip
&&{\dd^6 V_\eff\over \dd v^6} \Bigg|_{v=M^{1/2}} = \nu(M) = \nu 
\label{rencond}
\eeqn
The total effective potential to O($\hbar^2$) takes the form
\beeq
V_\eff(v;n) = {\Abar v^2\over 2} + {\Bbar v^4\over 4!} 
 + {\Cbar v^6\over 6!} 
   + {\Dbar v^6\over 6!} \ln {\mu^{2(3-n)} \nu v^4\over 4\pi\mu^2}  ~~.
\label{scalarPotential0}
\eneq
One need not know exact values of $\Abar$, $\Bbar$, 
and $\Cbar$ as they are absorbed in the definition of renormalized coupling
constants.  1-loop corrections are absent after renormalization.
2-loop contributions to $\Dbar$ are important.

The pole terms in $V_\eff^{(2)}$ have been cancelled by a counter
term:
\beeq
\delta\nu^{(2)}_{\rm pole}
= - 6! \, {\hbar^2\over 32\pi^2} \, 
 \Big[ 3 \Big({\nu\over 36} \Big)^2 +\Big( {\nu\over 60}\Big)^2  \Big]
 \, {1\over n-3} 
 = - {7\hbar^2 \nu^2\over 120\pi^2} \, {1\over n-3}
\label{pole1}
\eneq
The final expression for the effective potential is 
\beeq
V_\eff(v)^{\rm pure~scalar} = {1\over 6!} \, \nu(M) \, v^6
+ {1\over 6!} \, {7\hbar^2\over 120\pi^2} \, \nu(M)^2 \, v^6 
 \bigg( \ln {v^4\over M^2} - {49\over 5} \bigg) ~.
\label{scalarPotential1}
\eneq
The model is infrared free, the beta function being given by
\beeq
\beta(\nu) = \mu {\dd \nu(\mu) \over \dd \mu} 
= + {7\hbar^2\over 60\pi^2} \, \nu^2~~.
\label{beta-function}
\eneq

The logarithmic correction in (\ref{scalarPotential1}) 
 turns the tree-level minimum at $v=0$ into a maximum. 
However, it is premature to conclude that symmetry is spontaneously broken, 
as  the new minimum of the potential is located outside 
the domain of validity  of perturbation theory:
$\nu \ln v_{\rm min}^2/M \sim - 60\pi^2/7$. Higher order corrections
would drastically change the location of the minimum. 
 This also happens in 
four-dimensional pure scalar theory as was pointed out by Coleman and 
Weinberg.\cite{CW}
Our conclusion  is not in contradiction with the results of
other works \cite{MCS} because of  different renormalization 
conditions.

Now, let's turn to the case where the Maxwell-Chern-Simons gauge fields are
present, and adopt the full Lagrangian (\ref{Lagrangian1}). 
One loop corrections are easy to find.  Since 
\beeq
\det K = \xi^{-1} [ap^2 -(ev)^2]^{n-3} [p^2 - \xi(ev)^2]
[\{ ap^2 - (ev)^2 \}^2 - \kappa^2 \hp^2]~~, 
\eneq
one loop correction to the effective potential (at $n=3$) is
\beeq
V_\eff^{(1)}(v) =
 -{\hbar\over 12\pi} \bigg\{ m_1(v)^3 + m_2(v)^3
+   m_+(v)^3 +  m_-(v)^3  - [\xi (ev)^2]^{3/2} \bigg\}
\label{1loopV1}
\eneq
which, in the Landau gauge $\xi=0$,  becomes
\beeq
V_\eff^{(1)}(v) =
-{\hbar\over 12\pi} \Bigg\{ \bigg[ \Big({\nu\over 24}\Big)^{3/2}
+  \Big({\nu\over 120}\Big)^{3/2} \bigg] \, v^6 +
\sqrt{ {\kappa^2\over a^2}  + {4(ev)^2\over a}} ~
\Bigg( {\kappa^2\over a^2}  + {(ev)^2\over a} \Bigg) 
- {\kappa^3 \over a^3} \Bigg\} ~~.
\label{1loopV2}
\eneq
After renormalization the one-loop effective potential is 
\beqn
&&V_\eff(v)^{\rm 1~loop}
= {\nu\over 6!} \, v^6 + {\hbar\over 12\pi} {e^6\over a^3} \,F(x) \cr
\noalign{\kern 10pt}
&&F(x) = 3 \tkappa x^2 - (\tkappa^2 + 4x^2)^{1/2} (\tkappa^2 + x^2)
+ {2 \tkappa^4
(240 \tM^2 -62 \tM \tkappa^2 +  \tkappa^4)  \over
(4 \tM +  \tkappa^2 )^{11/2} } \, x^6 + \tkappa^3\cr
&&x =  {\sqrt{a} v\over e} ~~,~~ \tM = {a M\over e^2}
 ~~,~~ \tkappa = {\kappa\over e^2}~~.
\label{1loopV3}
\eeqn

If $\kappa \not= 0$, one can take $M=0$ at least at the 1 loop level.
With this choice $F(x)/\tkappa^3 = 3y^2 - (1+4y^2)^{1/2} (1+y^2) + 2y^6 +1$
where $y=x/\tkappa$.  $F(x) \ge 0$ so that $V_\eff^{\rm 1~loop}$ takes
the absolute minimum at $v=0$.  However, if one chooses $\tM = \tkappa^2$,
$F(x)/\tkappa^3 = 3y^2 - (1+4y^2)^{1/2} (1+y^2) + 0.00512 \, y^2 + 1$ which
has a minimum at a nonvanishing $x$.  In other words, spontaneous 
symmetry breaking may or may not occur,
depending on the choice of the renormalization point for the coupling
constant $\nu$.

If $\kappa=0$, one cannot take the renormalization condition
$V^{(4)}_\eff(0) =0$ in (\ref{rencond}) as the $|v|^3$ term is singular at
$v=0$.  If one takes $V^{(4)}_\eff(M'^{1/2}) =0$, then 
$V_\eff = (\nu/6!) (v^6 - 15 M' v^4) - (e^2/a)^{3/2} v^3/6\pi$.  The minimum
is located at $v\not= 0$.   Again,  the symmetry breaking seems to have 
resulted from the renormalization condition imposed.

In either case, $\kappa =0$ or $\kappa \not= 0$, one needs to go beyond one
loop.  We shall see below that two loop corrections are decisive in determining
the symmetry of the vacuum.  Nevertheless it is worthwhile to recognize that
the $a\go 0$ limit exists at one loop level.   In this limit, with
the renormalization condition (\ref{rencond}),
\beeq
\lim_{a\go 0} V_\eff(v)^{\rm 1-loop}  = {\nu\over 6!} \, v^6~~~.
\label{1loopV4}
\eneq
In other words, in the pure Chern-Simons theory all one loop corrections
are absorbed in the renormalization conditions.

Now we evaluate two loop corrections to the effective potential.
Computations are similar to the one described above for the pure
scalar theory.   With the use of the decomposition of the gauge field
propagator (\ref{Kinverse2}),  gauge field contributions
to diagrams of $\theta$ shape are reduced to the $I(m_1,m_2,m_3)$ function
defined in  (\ref{Ifunction}).  Contributions coming from the last term in
(\ref{Kinverse2}) vanish in the $n\go 3$ limit at the two loop level.

There are five types of diagrams.   

\noindent (1) Two scalar loops 
%\vskip 0.0 cm
%\hskip 3cm 
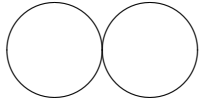
\begin{center}
\bpic(70,12)
\put(20,3){\circle{36}}
\put(56,3){\circle{36}}
\epic
\end{center}
\vskip 0.5 cm
\beeq
V_\eff^{(q1)} = {\hbar^2 \over {(4\pi)^2}} 
  \Bigg\{ 3\left({\lambda\over 4!} + {{15\nu v^2} \over {6!}} \right)m_1^2 
 + 3\left( {\lambda\over {4!}} + {{3\nu v^2}\over {6!}} \right) m_2^2 
 + 2\left({\lambda\over{4!}} + {{9\nu v^2}\over {6!}} \right)m_1 m_2 \Bigg\}~.
\label{2loopVQ1}
\eneq

\noindent (2) One scalar and one gauge loop
%\vskip 0. cm
%\hskip 2.5cm 
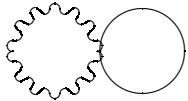
\begin{center}
\bpic(90,16)(0,13)  %(86, 16)(-10,13)
\put(32,32){\oval(4,4)[t]}
\put(36,32){\oval(4,4)[bl]}
\put(28,32){\oval(4,4)[br]}
\multiput(36,30)(0.2,0.1){4}{\line(1,0){0.1}}
\multiput(28,30)(-0.2,0.1){4}{\line(1,0){0.1}}
\multiput(36.8,30.4)(0.1,0.1){16}{\line(1,0){0.1}}
\multiput(27.2,30.4)(-0.1,0.1){16}{\line(1,0){0.1}}
\multiput(38.4,32.0)(0.2,0.1){4}{\line(1,0){0.1}}
\multiput(25.6,32.0)(-0.2,0.1){4}{\line(1,0){0.1}}
\put(39.2,30.4){\oval(4,4)[tr]}
\put(24.8,30.4){\oval(4,4)[tl]}
\multiput(41.2,30.4)(-0.1,-0.2){12}{\line(0,1){0.1}}
\multiput(22.8,30.4)(0.1,-0.2){12}{\line(0,1){0.1}}
\put(42.0,28.0){\oval(4,4)[bl]}
\put(22.0,28.0){\oval(4,4)[br]}
\multiput(42.0,26.0)(0.2,0.1){15}{\line(1,0){0.1}}
\multiput(22.0,26.0)(-0.2,0.1){15}{\line(1,0){0.1}}
\put(45.0,25.5){\oval(4,4)[tr]}
\put(19.0,25.5){\oval(4,4)[tl]}
\multiput(47.0,25.5)(-0.1,-0.2){4}{\line(0,1){0.1}}
\multiput(17.0,25.5)(0.1,-0.2){4}{\line(0,1){0.1}}
\multiput(46.6,24.7)(-0.1,-0.1){15}{\line(0,1){0.1}}
\multiput(17.4,24.7)(0.1,-0.1){15}{\line(0,1){0.1}}
\multiput(45.1,23.2)(-0.1,-0.2){7}{\line(0,1){0.1}}
\multiput(18.9,23.2)(0.1,-0.2){7}{\line(0,1){0.1}}
\put(46.5,22.0){\oval(4,4)[bl]}
\put(17.5,22.0){\oval(4,4)[br]}
\multiput(46.5,19.9)(0.1,0){15}{\line(1,0){0.1}}
\multiput(17.5,19.9)(-0.1,0){15}{\line(1,0){0.1}}
\put(48.0,18.0){\oval(4,4)[r]}
\put(16.0,18.0){\oval(4,4)[l]}

\put(32,0){\oval(4,4)[b]}
\put(36,0){\oval(4,4)[tl]}
\put(28,0){\oval(4,4)[tr]}
\multiput(36,2)(0.2,-0.1){4}{\line(1,0){0.1}}
\multiput(28,2)(-0.2,-0.1){4}{\line(1,0){0.1}}
\multiput(36.8,1.6)(0.1,-0.1){16}{\line(1,0){0.1}}
\multiput(27.2,1.6)(-0.1,-0.1){16}{\line(1,0){0.1}}
\multiput(38.4,0.0)(0.2,-0.1){4}{\line(1,0){0.1}}
\multiput(25.6,0.0)(-0.2,-0.1){4}{\line(1,0){0.1}}
\put(39.2,1.6){\oval(4,4)[br]}
\put(24.8,1.6){\oval(4,4)[bl]}
\multiput(41.2,1.6)(-0.1,0.2){12}{\line(0,1){0.1}}
\multiput(22.8,1.6)(0.1,0.2){12}{\line(0,1){0.1}}
\put(42.0,4.0){\oval(4,4)[tl]}
\put(22.0,4.0){\oval(4,4)[tr]}
\multiput(42.0,6.0)(0.2,-0.1){15}{\line(1,0){0.1}}
\multiput(22.0,6.0)(-0.2,-0.1){15}{\line(1,0){0.1}}
\put(45.0,6.5){\oval(4,4)[br]}
\put(19.0,6.5){\oval(4,4)[bl]}
\multiput(47.0,6.5)(-0.1,0.2){4}{\line(0,1){0.1}}
\multiput(17.0,6.5)(0.1,0.2){4}{\line(0,1){0.1}}
\multiput(46.6,7.3)(-0.1,0.1){15}{\line(0,1){0.1}}
\multiput(17.4,7.3)(0.1,0.1){15}{\line(0,1){0.1}}
\multiput(45.1,8.8)(-0.1,0.2){7}{\line(0,1){0.1}}
\multiput(18.9,8.8)(0.1,0.2){7}{\line(0,1){0.1}}
\put(46.5,10.0){\oval(4,4)[tl]}
\put(17.5,10.0){\oval(4,4)[tr]}
\multiput(46.5,12.1)(0.1,0){15}{\line(1,0){0.1}}
\multiput(17.5,12.1)(-0.1,0){15}{\line(1,0){0.1}}
\put(48.0,14.0){\oval(4,4)[r]}
\put(16.0,14.0){\oval(4,4)[l]}
\put(65,16){\oval(32,32)[b]}
%\put(65,16){\oval(32,32)[b]}
%\put(65,16){\oval(32,32)[t]}
\put(65,16){\oval(32,32)[t]}
\put(49,16){\circle*{1}}
\epic 
\end{center}
\vspace{0.5cm}
\beeq
V_{\eff}^{(q2)} = {e^2 \hbar^2 \over{16\pi^2}  a} 
{(m_1 + m_2) (m_+^2+m_-^2) \over m_+ + m_-} ~. 
\label{2loopVQ2}
\eneq

\noindent (3) $\theta$-shape diagram with pure scalar fields 
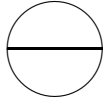
\begin{center}
\bpic(25, 0)
\put(18,3){\circle{36}}
\put(0,3){\line(1,0){36}}
\epic
\end{center}
\vskip 0.2 cm
\beqn
V_{\eff}^{(c1)} =  -{\hbar^2 \over {32\pi^2}} 
          \Bigg[3\bigg({\lambda\over 3!} v + {\nu \over 36}v^3\bigg)^2 
       + \bigg({\lambda\over 3!} v + {\nu \over 60}v^3\bigg)^2 \Bigg]
 \Bigg[-{1\over{n-3}} - \gamma_E + 1 + \ln {4 \pi} \Bigg] &&\cr
\myskip
      +{\hbar^2 \over {32\pi^2}}
         \bigg\{ 3 \bigg({\lambda\over 3!} v + {\nu \over 36}v^3\bigg)^2 
                      \ln {(3m_1)^2 \over \mu^2}
             + \bigg({\lambda\over 3!} v + {\nu \over 60}v^3\bigg)^2
 	              \ln {(m_1 + 2m_2)^2\over \mu^2} \bigg\} ~.&&
\label{2loopVC1}
\eeqn

\noindent (4) $\theta$-shape diagram with two scalar and one gauge propagators 
%\vskip 0. cm
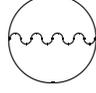
\begin{center}
\bpic(70,16)(0,13)
%\put(6,16){$\mu$}
%\put(71,16){$\nu$}
\multiput(26,19)(8,0){4}{\oval(4,4)[t]}
\multiput(30,19)(8,0){4}{\oval(4,4)[b]}
\put(40,19){\oval(33,33)[b]}
\put(40,19){\oval(33,33)[t]}
\epic
\end{center}
\vskip 0.0 cm 
\beqn
&&V_{\eff}^{(c2)}= {e^2 \hbar^2\over 64 \pi^2 a}   
\Bigg\{ \bigg[ 2(m_1^2 + m_2^2) - (m_+ + m_-)^2  + 3 m_3^2 \bigg] 
            \bigg[ -{1\over{n-3}} - \gamma_E + 1 + \ln {4\pi} \bigg] \cr
\myskip
&& \hskip 0.5 cm 
   + 2 \bigg [ m_1m_2 
   - {(m_1+m_2) \big\{ 2(m_1-m_2)^2 + m_+^2+m_-^2 \big\} \over m_+ + m_-}
       \bigg ] 
   - {(m_1^2 - m_2^2)^2 \over m_3^2} \ln {(m_1 + m_2)^2\over \mu^2}
   \cr
\myskip
&& \hskip 0.5 cm 
 - \sum_{a = \pm}  {2m_a^2(m_1^2 + m_2^2)   - m_a^4  - (m_1^2 - m_2^2)^2
      \over m_a(m_+ + m_-)} \ln{(m_a + m_1 + m_2)^2 \over \mu^2} 
 - {5 \over 6} {\kappa^2 \over a^2}
     \Bigg\} ~. 
\label{2loopVC2}
\eeqn

\noindent (5) $\theta$-shape diagram with two gauge and one scalar propagators
%\vskip 0. cm
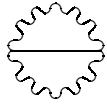
\begin{center}
\bpic(75,16)(0,13)
\put(32,32){\oval(4,4)[t]}
\put(36,32){\oval(4,4)[bl]}
\put(28,32){\oval(4,4)[br]}
\multiput(36,30)(0.2,0.1){4}{\line(1,0){0.1}}
\multiput(28,30)(-0.2,0.1){4}{\line(1,0){0.1}}
\multiput(36.8,30.4)(0.1,0.1){16}{\line(1,0){0.1}}
\multiput(27.2,30.4)(-0.1,0.1){16}{\line(1,0){0.1}}
\multiput(38.4,32.0)(0.2,0.1){4}{\line(1,0){0.1}}
\multiput(25.6,32.0)(-0.2,0.1){4}{\line(1,0){0.1}}
\put(39.2,30.4){\oval(4,4)[tr]}
\put(24.8,30.4){\oval(4,4)[tl]}
\multiput(41.2,30.4)(-0.1,-0.2){12}{\line(0,1){0.1}}
\multiput(22.8,30.4)(0.1,-0.2){12}{\line(0,1){0.1}}
\put(42.0,28.0){\oval(4,4)[bl]}
\put(22.0,28.0){\oval(4,4)[br]}
\multiput(42.0,26.0)(0.2,0.1){15}{\line(1,0){0.1}}
\multiput(22.0,26.0)(-0.2,0.1){15}{\line(1,0){0.1}}
\put(45.0,25.5){\oval(4,4)[tr]}
\put(19.0,25.5){\oval(4,4)[tl]}
\multiput(47.0,25.5)(-0.1,-0.2){4}{\line(0,1){0.1}}
\multiput(17.0,25.5)(0.1,-0.2){4}{\line(0,1){0.1}}
\multiput(46.6,24.7)(-0.1,-0.1){15}{\line(0,1){0.1}}
\multiput(17.4,24.7)(0.1,-0.1){15}{\line(0,1){0.1}}
\multiput(45.1,23.2)(-0.1,-0.2){7}{\line(0,1){0.1}}
\multiput(18.9,23.2)(0.1,-0.2){7}{\line(0,1){0.1}}
\put(46.5,22.0){\oval(4,4)[bl]}
\put(17.5,22.0){\oval(4,4)[br]}
\multiput(46.5,19.9)(0.1,0){15}{\line(1,0){0.1}}
\multiput(17.5,19.9)(-0.1,0){15}{\line(1,0){0.1}}
\put(48.0,18.0){\oval(4,4)[r]}
\put(16.0,18.0){\oval(4,4)[l]}

\put(32,0){\oval(4,4)[b]}
\put(36,0){\oval(4,4)[tl]}
\put(28,0){\oval(4,4)[tr]}
\multiput(36,2)(0.2,-0.1){4}{\line(1,0){0.1}}
\multiput(28,2)(-0.2,-0.1){4}{\line(1,0){0.1}}
\multiput(36.8,1.6)(0.1,-0.1){16}{\line(1,0){0.1}}
\multiput(27.2,1.6)(-0.1,-0.1){16}{\line(1,0){0.1}}
\multiput(38.4,0.0)(0.2,-0.1){4}{\line(1,0){0.1}}
\multiput(25.6,0.0)(-0.2,-0.1){4}{\line(1,0){0.1}}
\put(39.2,1.6){\oval(4,4)[br]}
\put(24.8,1.6){\oval(4,4)[bl]}
\multiput(41.2,1.6)(-0.1,0.2){12}{\line(0,1){0.1}}
\multiput(22.8,1.6)(0.1,0.2){12}{\line(0,1){0.1}}
\put(42.0,4.0){\oval(4,4)[tl]}
\put(22.0,4.0){\oval(4,4)[tr]}
\multiput(42.0,6.0)(0.2,-0.1){15}{\line(1,0){0.1}}
\multiput(22.0,6.0)(-0.2,-0.1){15}{\line(1,0){0.1}}
\put(45.0,6.5){\oval(4,4)[br]}
\put(19.0,6.5){\oval(4,4)[bl]}
\multiput(47.0,6.5)(-0.1,0.2){4}{\line(0,1){0.1}}
\multiput(17.0,6.5)(0.1,0.2){4}{\line(0,1){0.1}}
\multiput(46.6,7.3)(-0.1,0.1){15}{\line(0,1){0.1}}
\multiput(17.4,7.3)(0.1,0.1){15}{\line(0,1){0.1}}
\multiput(45.1,8.8)(-0.1,0.2){7}{\line(0,1){0.1}}
\multiput(18.9,8.8)(0.1,0.2){7}{\line(0,1){0.1}}
\put(46.5,10.0){\oval(4,4)[tl]}
\put(17.5,10.0){\oval(4,4)[tr]}
\multiput(46.5,12.1)(0.1,0){15}{\line(1,0){0.1}}
\multiput(17.5,12.1)(-0.1,0){15}{\line(1,0){0.1}}
\put(48.0,14.0){\oval(4,4)[r]}
\put(16.0,14.0){\oval(4,4)[l]}
\put(15,16){\line(1,0){34}}
%\put(15,16){\circle*{1}}
%\put(49,16){\circle*{1}}
\epic 
\end{center}
\vskip .2cm
\beqn
&&V_{\eff}^{(c3)} = - 
  {3\hbar^2 e^4 v^2 \over 64\pi^2a^2} 
\bigg[ -{1\over n-3} - \gamma_E   + 1 + \ln {4\pi} \bigg] \cr
\myskip
&&\hskip .3cm
 - {\hbar^2 e^4 v^2 \over 32\pi^2a^2} 
    \bigg[ - {2m_1 \over m_+ + m_-} -  {2m_1^2 + 12 m_3^2\over (m_+ + m_-)^2}
              + 3 \bigg] \cr
\myskip
&&\hskip .3cm  + {\hbar^2 e^4 v^2 \over 128\pi^2a^2} 
\Bigg\{  
    {2\big[(m_+ - m_-)^2 - m_1^2 \big]^2
       \over m_3^2(m_+ + m_-)^2}
              \ln {(m_+ + m_- + m_1)^2 \over \mu^2} 
    + {m_1^4\over m_3^4} \ln {m_1^2 \over \mu^2} \cr
\myskip
&& \hskip .3 cm
+ \sum_{a = \pm} \bigg[ {(4m_a^2 - m_1^2)^2 \over m_a^2(m_+ + m_-)^2}
              \ln {(2m_a + m_1)^2 \over \mu^2}
    - {2(m_a^2 - m_1^2)^2 \over m_3^2 m_a (m_+ + m_-)}
              \ln {(m_a + m_1)^2 \over \mu^2} \bigg] \Bigg\} . 
\label{2loopVC3}
\eeqn 
Notice that there appear logarithmic corrections at the two loop level, 
which give dominant contributions at small and large $v$.

Exact renormalization of the effective potential is straightforward but
tedious.   The main interest of the present paper is to know whether or 
not the symmetry is spontaneously broken in the massless scalar
electrodynamics.  It is therefore sufficient to look at the effective
potential at small and large $v$.
The behavior at small $v$ shows whether the tree
level minimum turns out to be a maximum or not,  while the behavior at large $v$
 tells us about the stability of the theory. 

Full examination of the effective potential will be reported in a separate
paper.  There one can determine the location of the minimum, and know
 whether quantum corrections are small near the minimum so that the use of
perturbation theory is justified.  The validity of perturbation theory
should give a condition on the parameters $\nu$, $e^2/\kappa$, and $a$.

\noindent $\underline{\hbox{Behaviour at small $v$ and $\kappa \not= 0$}}$

For small $v$ and $\kappa \not= 0$  
\beqn
&&m_1^2 \sim {\nu\over 24} v^4 ~~~,~~~ m_2^2 \sim {\nu\over 120} v^4 ~~~,\cr
\noalign{\kern 8pt}
&&m_3^2 \sim {e^2\over a} v^2 ~~~,~~~
m_+^2 \sim {\kappa^2\over a^2} ~~~,~~~ 
m_-^2 \sim {e^4\over \kappa^2} v^4  ~~~.
\label{Msmall}
\eeqn
One can  write the two loop corrections
to the effective potential in the  form 
\beeq
V_{\eff}^{(2)} =
 \sum_{n=1}^\infty \gamma_{2n}v^{2n}   
-  \big( \xi_2v^2 + \xi_4 v^4  + \xi_6 v^6 \big) \, {1\over n-3}
           + \chi_6v^6 \ln {v^4} ~~.
\label{Vsmall1}
\eneq
Explicit examination confirms that there appears no $\ln v$,  $v^2 \ln v$, or
$v^4 \ln v$ term so that one can maintain the renormalization conditions
(\ref{rencond}). $\gamma_2$, $\gamma_4$, 
$\xi_2$, $\xi_4$, and $\xi_6$ terms are irrelevant as they are
absorbed by renormalization.  The behaviour of $V_\eff$ at small $v$ is 
controled by the $\chi_6$ term.
After renormalization (\ref{rencond}) the dominant behaviour is given by
\beeq
V_{\eff} =   \chi_6v^6 
           \ln v^4   + \cdots \hskip 1cm {\rm for ~small~} v~.
\label{Vsmall2}
\eneq 

Inserting (\ref{Msmall}) into (\ref{2loopVC1}) - (\ref{2loopVC3}),  one
finds
\beeq
\chi_6 = {\hbar^2 \over {128\pi^2}} \Bigg\{ 
16 \Big( {e^2\over \kappa} \Big)^4 
- {11\over 30}   \Big( {e^2\over \kappa} \Big)^2 \, \nu
+{7\over 675} \, \nu^2 \Bigg\}  > 0~~. 
\vphantom{
\chi_6 = {\hbar^2 \over {128\pi^2}} \bigg\{ 16 \bigg(
           {e^4 \over \kappa^2} - {11\nu \over 960} \bigg)^2
          + \bigg({7\over 675} - {121\over 57600} \bigg)\nu^2 \bigg\}
}
\label{chi6}
\eneq
We have found that the tree level minimum at $v=0$ has turned into a maximum.

\noindent $\underline{\hbox{Behaviour at large $v$}}$

For large $v$ the masses are given by 
\beqn
&& m_1^2 \sim {\nu \over 24} v^4 ~~~,~~~
   m_2^2 \sim {\nu \over 120} v^4 \cr
\myskip
&& m_+^2 ,  m_-^2 , m_3^2 \sim {e^2 v^2 \over a} ~~~.
\label{Mlarge}
\eeqn
We suppose that $a\not= 0$. 
Two loop corrections to the effective potential for large $v$ are summarized as
\beeq
V_{\eff}^{(2)} = \sum_{n=0}^\infty \omega_{6-2n} v^{6- 2n}  +
\big( \xi_2v^2 + \xi_4 v^4    + \xi_6 v^6 \big)
\bigg( -{1\over n-3} +  \ln v^4 \bigg) ~.
\label{Vlarge1}
\eneq
The logarithm and pole terms have the same coefficients for large $v$, as all
the logarithm terms in 2 loop corrections (\ref{2loopVC1}) - (\ref{2loopVC3})
contain
$m_1/\mu$ or
$m_2/\mu$ in their arguments, and $\mu^2$ originally appears in the combination
of $- (n-3)^{-1} +\ln 4\pi\mu^2 $. 

After renormalization
\beeq
V_{\eff} =   \xi_6 v^6 
           \ln v^4  + \cdots \hskip 1cm {\rm for ~large~} v~.
\label{Vlarge2}
\eneq 
The $\xi_6$ term controles the behaviour at large $v$.
Insertion of (\ref{Mlarge}) into 
(\ref{2loopVC1}) - (\ref{2loopVC3}) yields 
\beeq
\xi_6 = {7\hbar^2 \over 120\pi^2} {\nu^2 \over 6!} >0 
\label{xi6}
\eneq              
which establishes the stability of the theory at two loop.  Combining the 
result (\ref{xi6}) with (\ref{chi6}), we conclude that symmetry is 
spontaneously broken by radiative corrections at two loop when $\kappa\not=0$.
(Of course, as mentioned above, we need to check that the minimum of the 
potential is located in the domain where perturbation theory is valid.)
The result is independent of the Maxwell coefficient $a$.

There are a couple of features to be recognized.  First, both $\chi_6$ and
$\xi_6$ are indenpendent of $a$.  Secondly $\xi_6$ is independent of $e$
and $\kappa$.
We have demonstrated these features at two loop by explicit evaluations.
Indeed, these properties remain true to all orders in perturbation theory.

To see it, we note that one of the parameters  $\kappa$, $e$, and $a$ is 
redundant.  So long as $a\not=0$, one can absorb $a$ in the redefinition of 
$A_\mu$.  The model with ($\kappa, e,a$) is equivalent to the model with
$(\kappa', e',a')=(\kappa/a, e/\sqrt{a}, 1)$.

$\xi_6$ is a dimensionless quantity so that it must be a function of 
$\nu$ and $\kappa'/e'^2$.  Since the $\xi_6$ term is associated with the 
large $v$ behaviour of the effective potential, one can make use of 
(\ref{Mlarge}) in evaluating each diagram.  However, none of $m_a(v)$
in (\ref{Mlarge}) contains $\kappa'$, and therefore $\xi_6$ must be a function
of $\nu$ only.  We have proven:

\noindent{\bf Theorem 1}  ~
{\sl $\xi_6$ defined in the behaviour of
the effective  potential at large $v$, (\ref{Vlarge2}), is independent of $a$,
$e$, and $\kappa$ to all orders in perturbation theory.}

Similarly $\chi_6$ is dimensionless, and is a function of $\nu$ and 
$\kappa'/e'^2=\kappa/e^2$.  It follows

\noindent {\bf Theorem 2} ~
{\sl $\chi_6$ defined in the behaviour
of the effective  potential at small $v$, (\ref{Vsmall2}), is independent of
$a$ to all orders in perturbation theory.} 

The existence of the limit of the vanishing Maxwell term $(a\go 0$) is by no
means obvious.  We have  shown that at the two loop level the $a\go 0$
limit exists  at both small and large $v$.  This supports the idea of improving
 high momentum behavior of the gauge field propagator in Chern-Simons theory by
adding  a Maxwell term.\cite{Martin,FPS}

$\xi_6$ is directly related to the beta function for the coupling $\nu$ and 
a similar theorem follows for the beta function itself.  The Lagrangian 
(\ref{Lagrangian1}) defines various vertices in perturbation theory.  Let
$V_4$, $V_6$, $V_{3A}$, $V_{4A}$, and $V_{3F}$ denote the numbers of vertices
$\phi^4$, $\phi^6$, $A\phi \dd \phi$, $A^2 \phi^2$, and $\phi c^\dagger c$
in a given Feynman diagram $F$, respectively.  $E$ denotes the number of
external lines.  If $a \not= 0$, all  propagators including gauge boson
propagators behave as $1/p^2$ for large $p^2$.    The superficial degree of
divergence by power counting is then given by
\beeq
\omega(F) = 3 - {E\over 2} - V_4 - V_{4A} 
  - {1\over 2} V_{3A} - {3\over 2} V_{3F} ~~.
\label{divergence-degree}
\eneq

For  propagators, $\omega=2$ only if 
$V_4$=$V_{3A}$=$V_{4A}$=$V_{3F}$=0.  That is, the wave funcion renormalization
constant $Z_\phi$ for $\phi$ field gets divergent contributions (or  pole
terms in the dimensional  regularization scheme) from diagrams which contain
only
$\phi^6$ vertices.  Hence $Z_\phi$ in the minimal subtraction scheme depends
solely on $\nu$, being  independent of the gauge couplings.  
Similarly, for graphs contributing to the $\phi^6$ coupling ($E=6$),
$\omega=0$ only if $V_4$=$V_{3A}$=$V_{4A}$=$V_{3F}$=0 so that $Z_\nu$
depends on $\nu$ only.  Hence we have:

\noindent {\bf Theorem 3} ~ 
{\sl The beta function for $\nu$ is independent of gauge couplings.}

\noindent
In particular, $\beta_\nu$  at two loop is given by
(\ref{beta-function}). 

We have calculated the effective potential in the Maxwell-Chern-Simons 
theory of massless scalar fields ($m=\lambda=0$) 
up to two loop corrections in the Landau gauge. 
At the two loop level, the $U(1)$ symmetry is spontaneously broken for $\kappa
\not= 0$. A closer investigation has revealed that
 the beta function for the $\phi^6$ scalar coupling is independent of the
gauge coupling.

We plan to give a full account of the two loop effective potential 
in a forthcoming paper.  We shall determine the location
of the absolute minimum of the potential, and the dependence of the potential
on the parameters $\nu$, $a$, $e$ and $\kappa$. In particular, 
in the $a\go 0$ limit
the requirement of the validity of perturbation theory around the minimum 
should yield a relation between $\nu$ and $\kappa/e^2$.  The initial parameters
($\nu, \kappa/e^2$) are replaced by ($\la \phi \ra, \kappa/e^2$), and we will
see more explicitly that the dimensional transmutation takes place.

\vskip 10pt

\baselineskip=16pt  
{\small
\leftline{\bf Acknowledgement}
This work was supported in part  by the U.S.\ Department of Energy
under contracts DE-AC02-83ER-40105.
}
\vskip 15pt

\leftline{\bf References}

\renewenvironment{thebibliography}[1]
	{\begin{list}{[$\,$\arabic{enumi}$\,$]}  % {\arabic{enumi}.}
	{\usecounter{enumi}\setlength{\parsep}{0pt}
	 \setlength{\itemsep}{0pt}  \renewcommand{\baselinestretch}{1.2}
         \settowidth
	{\labelwidth}{#1 ~ ~}\sloppy}}{\end{list}}

\end{document}